\documentclass[11pt]{article}
\usepackage{moriond,epsfig}

\def\Journal#1#2#3#4{{#1} {\bf #2}, #3 (#4)}

\def\NPB{{\em Nucl. Phys.} B}
\def\PLB{{\em Phys. Lett.}  B}
\def\PRL{\em Phys. Rev. Lett.}
\def\PRD{{\em Phys. Rev.} D}

\def\EPJ{{\em Eur. Phys. J.} C}

\def\be{\begin{equation}}
\def\ee{\end{equation}}
\def\bea{\begin{eqnarray}}
\def\eea{\end{eqnarray}}

\begin{document}
\vspace*{4cm}
\title{LEP 2 $e^+e^- \rightarrow f\overline{f}, \, \gamma\gamma (\gamma)$: results and interpretations}

\author{ Guy Wilkinson }

\address{Subdepartment of Particle Physics, University of Oxford, 
Denys Wilkinson Building, Keble Road, Oxford OX1 3RH, United Kingdom.}

\maketitle\abstracts{Results on LEP 2 
$e^+e^- \rightarrow f\overline{f}$ and $\gamma\gamma (\gamma)$ production
are presented.  These are compared with Standard Model predictions, and
then used to set limits on various New Physics models, including
Low Scale Gravity.  Finally the status of the LEP beam energy determination
from radiative return events is summarised.}

\section{Introduction}

\begin{figure}
\begin{center}
\epsfig{file=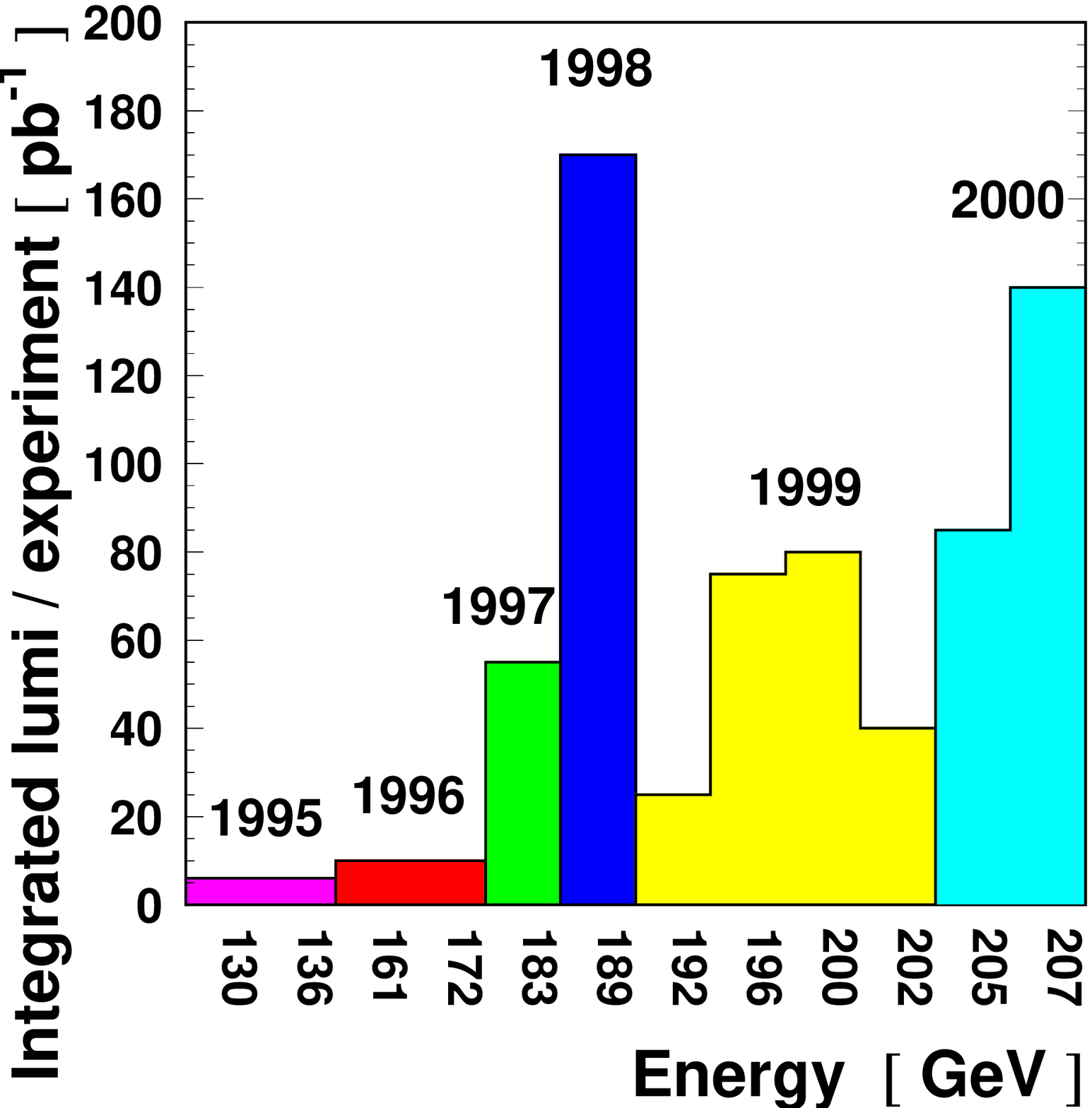,width=0.42\textwidth}
\epsfig{file=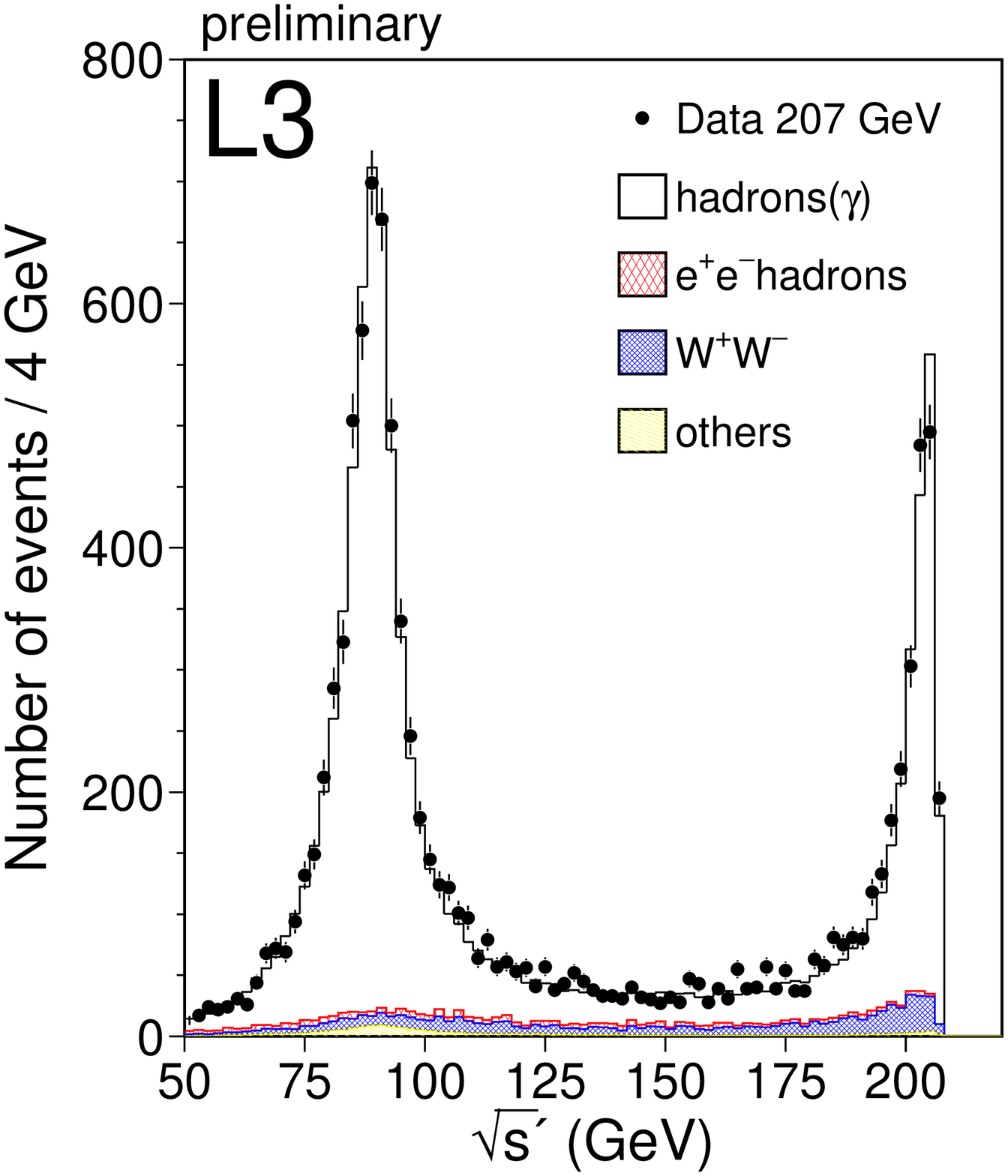,width=0.32\textwidth}
\end{center}
\caption[]{Left: approximate integrated luminosity at LEP 2.
Right: example $\sqrt{s^\prime}$ distribution in the $q \overline{q}$ channel.}
\label{fig:lumi}
\end{figure}

During the LEP 2 programme of 1995--2000 approximately 
$700 \,{\rm pb^{-1}}$ per experiment were collected 
at $e^+e^-$ centre-of-mass energies between 130 and 207 GeV.
How this integrated luminosity was distributed is indicated in 
figure~\ref{fig:lumi}.  As well as allowing precise studies
of 4 fermion final states, and the search for the Higgs boson
and supersymmetry, these data have been exploited in the
examination of 2 fermion and hard gamma production, and it
is this work which is reported here.  A collection of final
and preliminary results are presented, and these are interpreted 
in the context of the Standard Model and alternative theories.

\subsection{Selection}

The topologies of the $q \overline{q}$, $e^+ e^-$, 
$\mu^+ \mu^-$, $\tau^+ \tau^-$ and $\gamma \gamma \, (\gamma)$
final states are sufficiently distinctive for samples to be
selected with high efficiency and purity.
With these samples
cross-sections,  forward-backward asymmetries, and differential
cross-sections have been measured.   Within the $q \overline{q}$
sample, lifetime information from vertex detectors, together with 
other discriminating variables, has been used to isolate
$b \overline{b}$ and $c \overline{c}$ events.  These have been
used to measure the heavy quark branching ratios $R_{b,c} = 
\sigma_{b\overline{b}, \, c\overline{c}}/\sigma_{q\overline{q}}$ 
and corresponding forward-backward
asymmetries, $A^{b,c}_{\rm FB}$.

\subsection{$s^\prime$ reconstruction}

At LEP 2 energies the emission of initial photons is very probable.
Therefore the variable $\sqrt{s^\prime}$ is defined, which is the
centre-of-mass energy of the $e^+e^-$ system after initial state radiation.
Experimentally this is reconstructed from the direction of the
final state fermions, under the assumption that a single photon was
emitted undetected along the beam pipe.   
More sophisticated treatments allow for 
the emission of multiple photons and account for any observed radiation.
The resulting spectrum in $\sqrt{s^\prime}$ is shown in figure~\ref{fig:lumi}
for the L3 $q\overline{q}$ analysis.  Two clear peaks are visible:  that
at $\sqrt{s^\prime} \sim \rm m_Z$ 
comes from the so-called {\it radiative return}  events, whereas that
at $\sqrt{s^\prime} \sim \sqrt{s}$ consists of high energy, 
{\it non-radiative} events.  It is the non-radiative events which
are of most interest in testing the Standard Model and other theories.
In contrast, the radiative return events are used as
a calibration tool, as explained in section~\ref{sec:radret}

\subsection{LEP wide combination}

To achieve the best possible precision, efforts have been made
to average the published and preliminary results of the 
4 collaborations.  This has been done for all the avaliable 
$\mu^+\mu^-$, $\tau^+\tau^-$, $q \overline{q}$ and heavy quark cross-section
and asymmetry measurements. In performing this average careful attention has 
to be paid to correlated systematics, and to 
ensuring that the signal definition
is the same between the experiments.  (The latter requirement is not
always satisfied, for instance in the definition of non-radiative samples
or $\sqrt{s^\prime}$.  In this case appropriate 
corrections have been applied prior to combination.)

Table~\ref{tab:errors} shows the precision now achieved for the three
most important cross-section measurements, expressed in terms of 
deviation from the Standard Model prediction.  The total
experimental uncertainty is given, and the corresponding uncertainty coming
from theory.  The fact that the theoretical uncertainties are small
is the result of much effort invested at the time of the LEP 2 Monte
Carlo workshop~\cite{LEPMC}.   At present the corresponding
uncertainties in the $e^+e^-$ channel are more significant at $\sim 2\%$,
and therefore no LEP wide combination as yet been performed.
The total uncertainty on the asymmetries are dominated by statistics,
and are $0.012$ for the muons and $0.015$ for the taus. 

\begin{table}[t]
\caption{Approximate LEP wide precision on 2-fermion cross-sections.}
\vspace{0.4cm}
\begin{center}
\begin{tabular}{|l|c|c|c|}
\hline
Uncertainty  & $q\overline{q}$ & $\mu^+ \mu^-$  & $\tau^+ \tau^-$ \\ \hline
Experimental & 1.0 \% & 1.6 \% & 2.2 \% \\
Theoretical  & 0.3 \% & 0.4 \% & 0.4 \% \\ \hline
\end{tabular}
\end{center}
\label{tab:errors}
\end{table}  

Similar combinations are being made of the differential cross-sections, 
although this work is at a preliminary stage.  LEP wide averages exist for
the heavy flavour results and the $\gamma \gamma$ ($\gamma$) channel.
For the former it is the $b\overline{b}$ channel which is presently most
precise with a relative uncertainty of $2.5\%$ on $R_b$, averaged over 
all energies, and 0.06 on the asymmetry.   The   $\gamma \gamma$ ($\gamma$)
cross-section is precise to $1.2 \%$.

\section{Results and Comparison with Standard Model Predictions}

\begin{figure}
\begin{center}
\epsfig{file=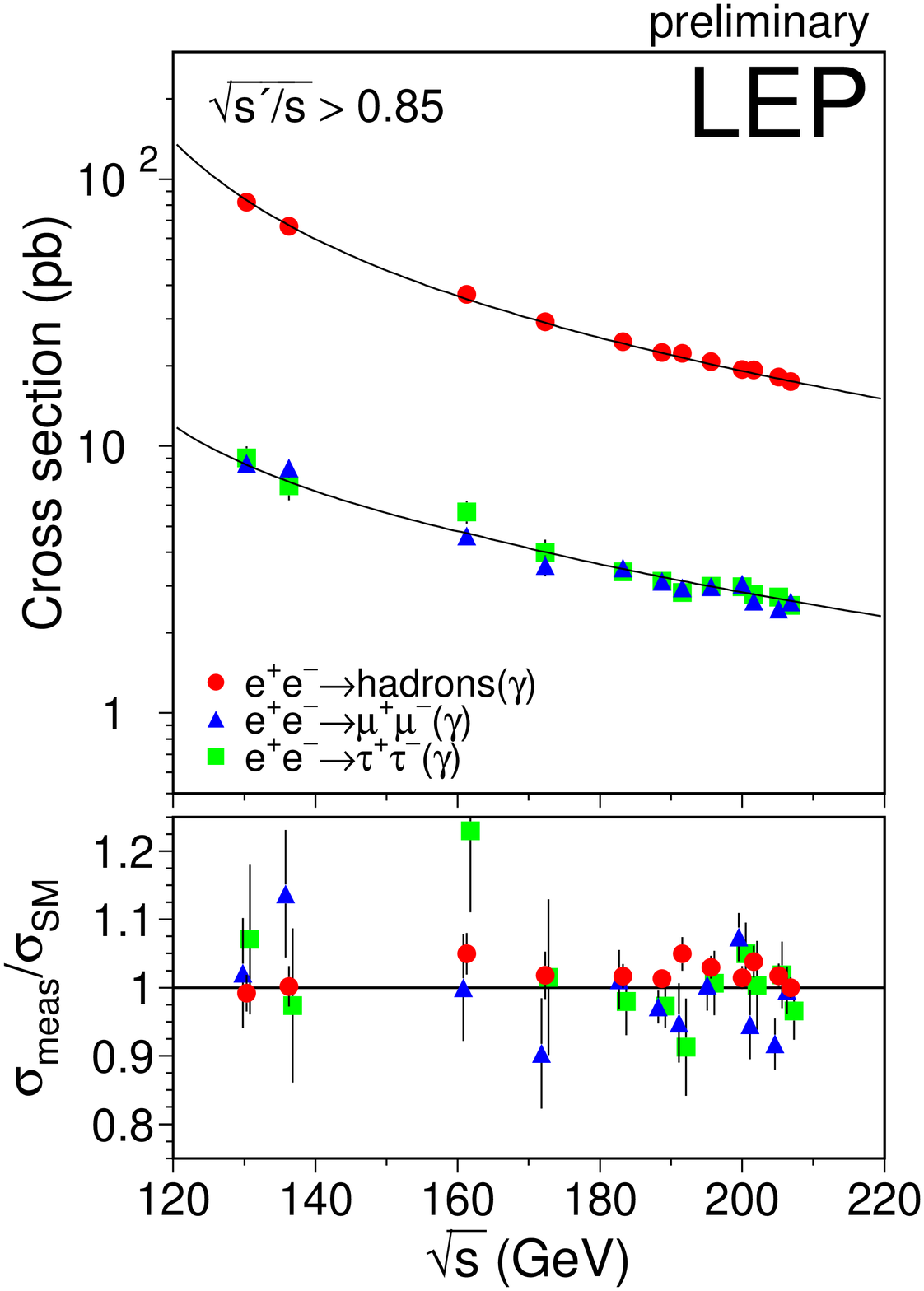,width=0.42\textwidth}
\epsfig{file=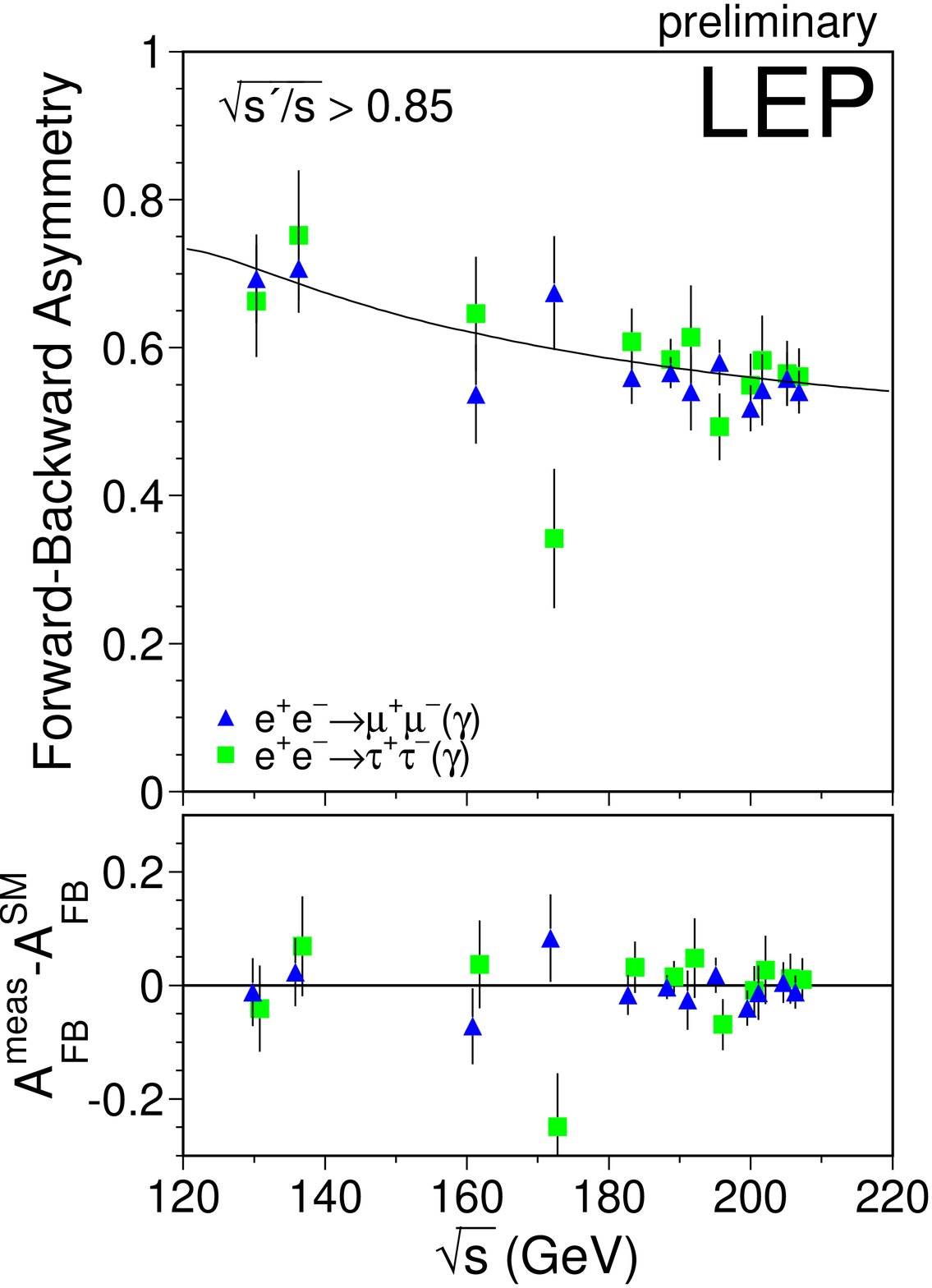,width=0.42\textwidth}
\end{center}
\caption[]{Cross-section results for the channels $e^+e^- 
\rightarrow q\overline{q} \,, \, \mu^+\mu^-$ and 
$\tau^+\tau^-$, and forward-backward asymmetries for
$e^+e^- \rightarrow  \mu^+\mu^-$ and $\tau^+\tau^-$, all with 
$\sqrt{s^\prime / s} > 0.85$.}
\label{fig:ffresults}
\end{figure}

Figure~\ref{fig:ffresults} show the LEP averaged cross-section and
forward backward asymmetry results for the non-radiative samples
($\sqrt{s^\prime / s} > 0.85$) as a function of $\sqrt{s}$, together
with the Standard Model prediction from ZFITTER~\cite{ZFITTER}.
Published and preliminary results  from all experiments and all energy
points are included~\cite{FFRESULTS}.
The agreement is generally good, although the $q\overline{q}$ cross-sections
are on average almost 2 sigma higher than expectation.

\begin{figure}
\begin{center}
\epsfig{file=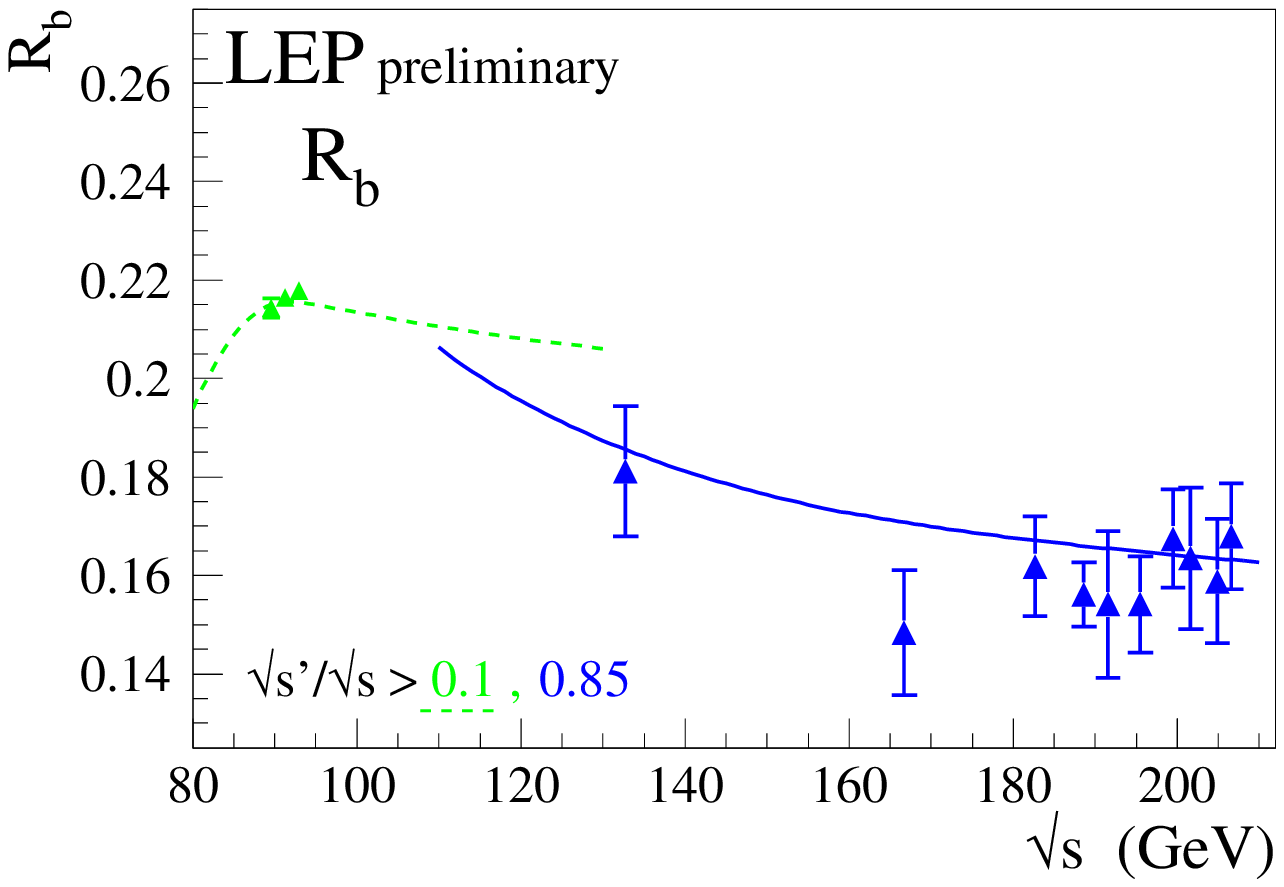,width=0.45\textwidth}
\epsfig{file=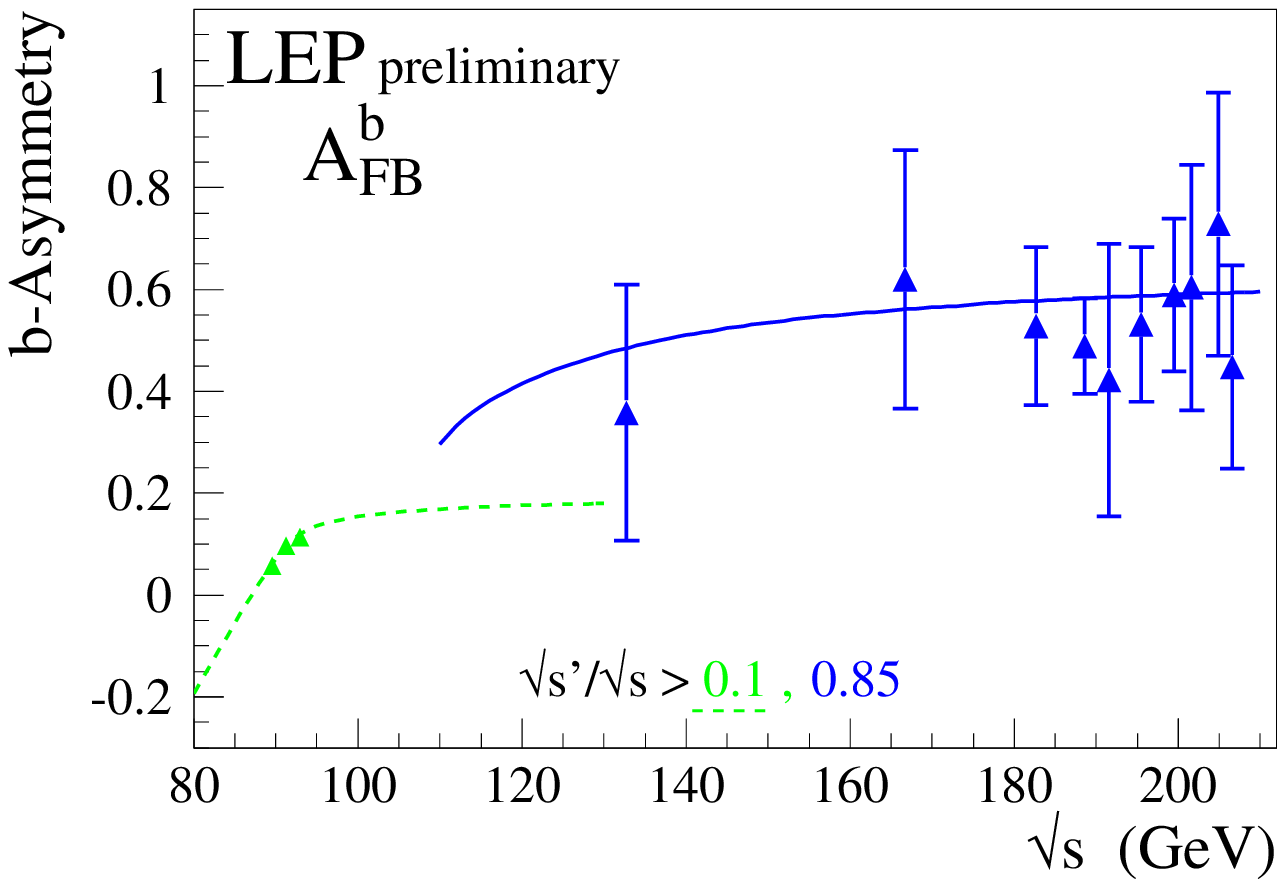,width=0.45\textwidth}
\end{center}
\caption[]{LEP combined $R_b$ and $A^b_{\rm FB}$ results against 
energy.}
\label{fig:bbresults}
\end{figure}

Figure~\ref{fig:bbresults} shows the LEP combined results on $R_b$ and
$A^b_{\rm FB}$.  These include published and preliminary measurements from
all experiments and energy points~\cite{FFRESULTS}, apart from ALEPH 2000 and OPAL 1999 and
2000, where no results have yet been announced. 
Again there is satisfactory agreement with expectation,
but with a tendency for the data to prefer a slightly lower $R_b$.

\begin{figure}
\begin{center}
\epsfig{file=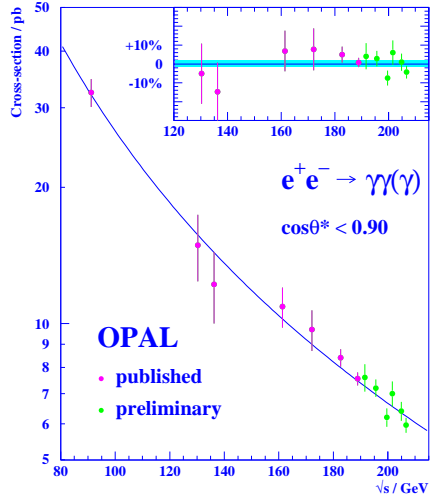,width=0.40\textwidth}  
\end{center}
\caption[]{OPAL measurements of the cross-section for 
$e^+e^- \rightarrow \gamma \gamma (\gamma)$, against $\sqrt{s}$. }
\label{fig:ggopal}
\end{figure}

Cross-section results from all significant data sets and experiments have been
combined for $e^+ e^- \rightarrow \gamma \gamma (\gamma)$~\cite{GGRESULTS}. When averaged
over energy these give a result of 
$\sigma_{\rm meas} / \sigma_{\rm QED} = 0.982 \pm 0.012$, where 
$\sigma_{\rm QED}$ is the theorerical prediction, which is known with
a precision of $1\%$. Figure~\ref{fig:ggopal} shows the results of OPAL
alone, and their dependence on energy. 

\section{Indirect Searches for New Physics}

Having first established the consistency of the fermion pair and 
$\gamma \gamma (\gamma)$ results with the Standard Model,  it is then
natural to use the same results to establish limits on New Physics.
Here three possibilities are explored.

\subsection{$Z^\prime$ bosons}

Many Standard Model extensions predict the existence of additional massive
neutral bosons, which are generically known as $Z^\prime$s.  These bosons
have a mass $M_{Z^\prime}$ and mix with the $Z^0$ with
a mixing angle $\theta_{ZZ^\prime}$.  The $Z^\prime$s' coupling to fermions
vary depending on the model in which they arise, but in general
have the potential to modify the cross-sections and asymmetries at LEP
energies.

The LEP 2 two fermion results have 
been used to set limits on  $Z^\prime$ bosons.
It turns out that the LEP 1 data are more sensitive to the mixing angle
and constrain this to be very small for all models.  Therefore here 
$\theta_{ZZ^\prime}$ is set to 0, and the LEP 2 data used to place limits 
on $M_{Z^\prime}$.   The results at 95\% confidence level are shown in
table~\ref{tab:zprime}, for a variety of string inspired models,
and the Sequential Standard Model (SSM).  In the latter the couplings
are the same as in the Standard Model, and it is  for this model that the
best limits are obtained.

\begin{table}[t]
\caption{LEP limits on the mass of $Z^\prime$ bosons.}
\vspace{0.4cm}
\begin{center}
\begin{tabular}{|l|c|c|c|c|c|}
\hline
Model & $\chi$ & $\psi$ & $\eta$ & LR & SSM \\ \hline
$M_{Z^\prime} >$ [GeV] & 678 & 463 & 436 & 800 & 1890 \\ \hline
\end{tabular}
\end{center}
\label{tab:zprime}
\end{table}  

\subsection{Contact Interactions}

Four fermion contact interactions parameterise New Physics, such as
heavy particle exchange or compositeness, in terms of an effective Lagrangian:

\begin{eqnarray}
{\cal{L_{\rm eff}}} & = & 
\frac{g^2}{(1 + \delta) \Lambda^2} \sum_{i,j = L,R} 
\eta_{ij} \overline{e_i} \gamma_\mu e_i \overline{f_j} \gamma^\mu f_j,
\nonumber
\end{eqnarray}

\noindent where $\delta = 1$ when $f=e$ and $0$ when $f \ne e$.  $\Lambda$
represents the characteristic energy scale of this New Physics.  Its
coupling strength is unknown, so by convention $g$ is chosen
such that $g^2/4\pi = 1$.

\begin{figure}
\begin{center}
\epsfig{file=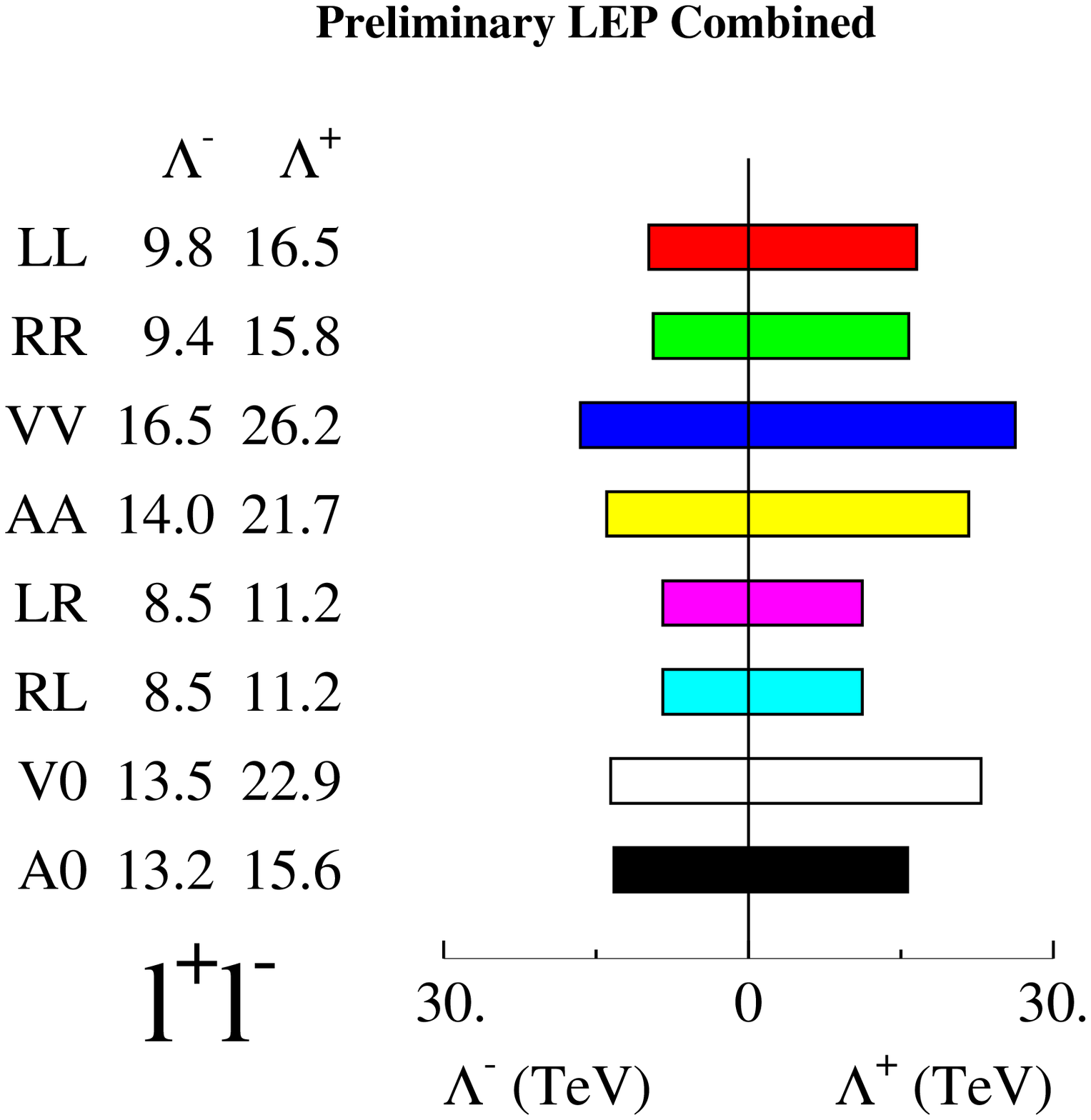,width=0.42\textwidth}  
\epsfig{file=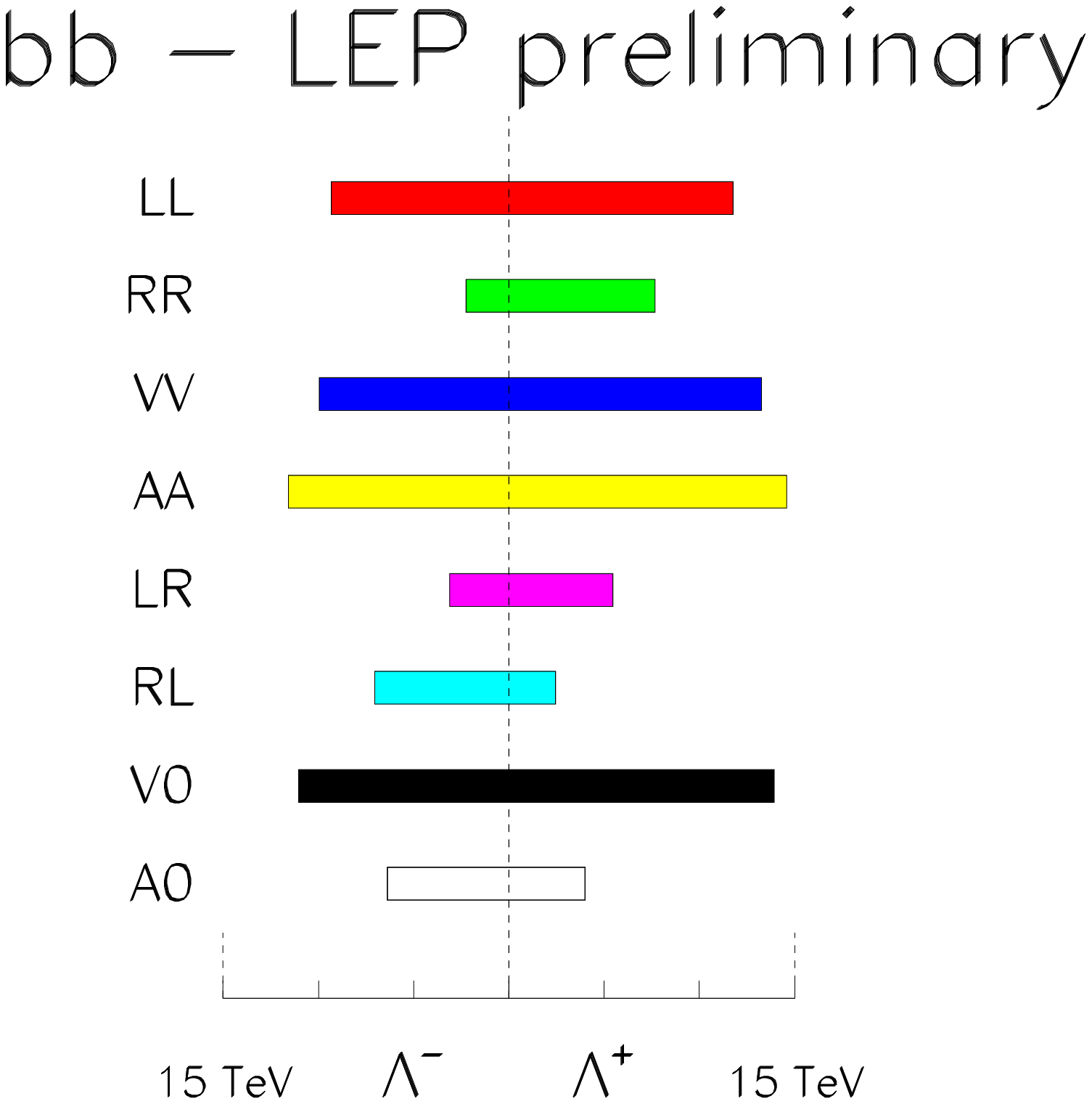,width=0.42\textwidth}  
\end{center}
\caption[]{$95\%$ limits on the contact interaction New Physics scale $\Lambda$ 
for various helicity structures, for leptons (left) and b quarks  (right).}
\label{fig:contact}
\end{figure}

Fits to the $e^+e^- \rightarrow l^+l^-$ data have been performed 
with $\epsilon = 1/\Lambda^2$ as a free parameter, so that $\epsilon = 0$
represents the limit of no new physics.  Models have been considered
with different helicity structures (LL,RR etc).   No evidence of
physics beyond the standard model has been found, and limits have
been obtained on $\Lambda$ of between 8.5 and 26.2 TeV, as shown in 
figure~\ref{fig:contact}.   The figure also includes the corresponding
limits for contact interactions between electrons and b quarks.
The superscript on $\Lambda$ signifies constructive (+) or
destructive (-) interference with the Standard Model.

\subsection{Low Scale Gravity (LSG)}
\label{sec:xtradim}

A long standing problem in physics is the huge difference in magnitude
between the electroweak scale ($M_{\rm ew} \sim 10^2$ GeV) and
that of gravity ($M_{\rm Planck} \sim 10^{19}$ GeV).   
Various possible solutions have been 
advanced, most notably SUSY. Recent proposals~\cite{LSG}, however, 
suggest
that the scale of the  electroweak and gravitational interactions
are in fact similar, but the latter appears diluted due to
graviton exchange occuring in more dimensions that the 4 
in which the Standard Model
particles propagate.   In this scheme the true gravity scale $M_D$  
($\sim M_{\rm ew}$) is related to its apparent scale by

\begin{eqnarray}
M^2_{\rm Planck} & = & (M_D)^{2+n} \, R^n.
\nonumber 
\end{eqnarray}

\noindent Here $n$ is the number of extra dimension, and $R$ is their
characteristic size.   It can be seen that assuming if, for instance,
$n=2$, then the radius of the compactified new dimensions will be $0.1 \, \rm mm$,
which is large compared with the predictions of more orthodox string theories.

If correct, this proposal makes plausible the possibility that 
LEP 2 two fermion production receive a contribution from
virtual graviton exchange.   In this case the differential
cross-section would assume the form:

\begin{eqnarray}
\frac{d \sigma}{d \cos \theta} & = & 
A(\cos \theta) \,+\, B(\cos \theta) \left[ \frac{\lambda}{M_S^4} \right]
               \,+\, C(\cos \theta) \left[ \frac{\lambda}{M_S^4} \right]^2,
\nonumber
\end{eqnarray}

\noindent where the $A(\cos \theta)$ is the Standard Model contribution,
the $ B(\cos \theta)$ term represents graviton--Standard Model
interference, and the $ C(\cos \theta)$ term represents pure graviton exchange.
$M_S$ ($\sim M_D$) is the cutoff energy for LSG, and $\lambda$ are other
possible model dependencies, which are chosen to be $\pm 1$ to allow for different
signs of interference.

\begin{figure}
\begin{center}
\epsfig{file=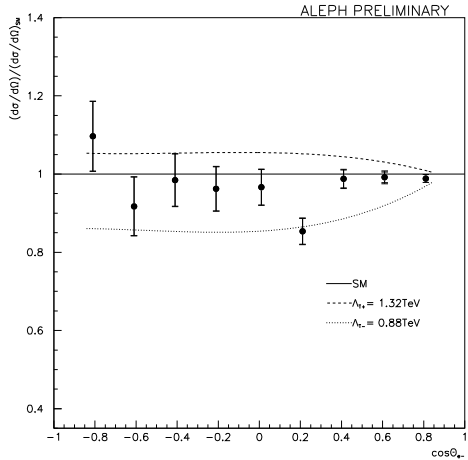,width=0.32\textwidth}  
\end{center}
\caption[]{ALEPH $e^+e^-$ differential cross-sction, normalised to the Standard
Model prediction.
Superimposed are the deviations expected for the fitted 95\% CL 
lower limits on $\Lambda_s$ ($\equiv \, (\pi/2)^{0.25} \times M_s$).}
\label{fig:alel}
\end{figure}

The experiments have interpreted their two fermion data in this context.
The best sensitivity is available from the $e^+e^-$ channel,
because of interference between LSG and the Standard Model t-channel diagram.
Figure~\ref{fig:alel} shows the observed ALEPH $e^+e^-$  differential
cross-section, normalised to the pure Standard Model expectation.  
The data are consistent with this description.  Superimposed are the deviations
resulting from the 95\% CL limit on $M_S$, which are 1.18~TeV and 0.81~TeV
for $\lambda =+1$ and $\lambda =-1$ respectively.

LSG can also affect boson pair production, where it is mainly seen in a modification
of the total cross-section.  A LEP combined analysis of the $\gamma \gamma (\gamma)$
events sets a limit of $M_s^{\lambda=+1} > 0.97 \, \rm TeV$ and
$M_s^{\lambda=-1} > 0.94 \, \rm TeV$~\cite{GGRESULTS}.   
Individual experiments have included other channels in 
combined analyses~\cite{ALLGRAV}.
Work is still required, however,
to produce a LEP combined limit.

\section{Determination of the LEP Beam Energy through Radiative Returns}
\label{sec:radret}

An important systematic error in the LEP $m_W$ measurement~\cite{CHRIS}
is the uncertainty in the LEP beam energy, $E_b$.  $E_b$ is presently
determined through machine based analyses~\cite{ECAL}, and is known
to a precision of $\sim 20 \, \rm MeV$.  The two fermion 
radiative return events provide a complementary way for the 
experiments to cross-check this number with their own data.
As the mass of the Z is extremely well known from LEP 1, 
the position of the radiative-return peak can be used to
determine $E_b$, and compared with the estimate of the energy
model used in the $m_W$ analysis.   In practice, however, the
analysis is extremely delicate, requiring in the hadronic channel
excellent knowledge of the jet reconstruction in the forward
region, where the events are generally found, and for all
channels good understanding of the initial and final 
state radiation processes.

\begin{figure}
\begin{center}
\epsfig{file=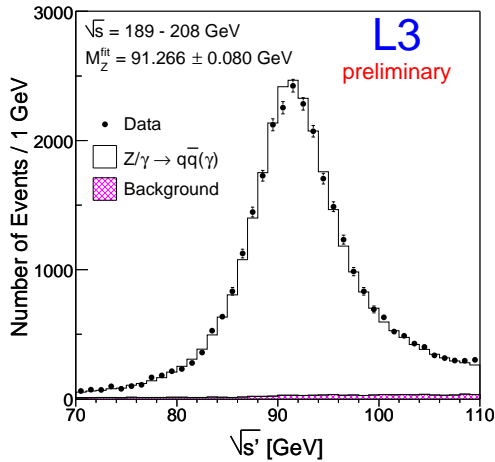,width=0.42\textwidth}  
\end{center}
\caption[]{L3 fitted $q\overline{q}$ radiative return peak for use in the 
$E_b$ determination.}
\label{fig:l3peak}
\end{figure}

Figure~\ref{fig:l3peak} shows the fitted radiative return  peak 
from a new L3 $q\overline{q}$ analysis.  Table~\ref{fig:radret}
gives a summary of available results~\cite{RADRET}, where $\Delta E_b$ is
the radiative return estimate of $E_b$ minus that coming
from the standard procedure.   No combination is
yet available taking account of the significant 
correlated systematics.   It can be seen that there is no
evidence of a disagreement within the accuracy of the measurements.

\begin{table}[t]
\caption{$E_b$ determination from radiative returns, compared with the machine estimate.}
\vspace{0.4cm}
\begin{center}
\begin{tabular}{|l|c|}
\hline
Measurement & $\Delta E_b$ [MeV] \\ \hline
ALEPH '97 $q\overline{q}$    & $-76 \pm 103$ \\
OPAL all                     & $-31 \pm 54$ \\
DELPHI $\mu^+\mu^-$          & $+76 \pm 96$ \\
L3 $\ge$ '98 $q\overline{q}$ & $-83 \pm 84$ \\ \hline
\end{tabular}
\end{center}
\label{fig:radret}
\end{table}

\section{Conclusions}

The high integrated luminosity delivered by LEP 2 at energies up to and beyond
200 GeV have allowed a precise study of two fermion and hard photon production,
and a detailed comparison with Standard Model predictions.  No significant
deviation has been found.    The same results have been used to set limits
on various New Physics models, including $Z^\prime$s, contact interactions
and Low Scale Gravity.   

A significant amount of work is required to finalise these studies.
Final publications are expected from each of the experiments, and these
results must then be correctly averaged.  There is scope for improvements
in the theoretical understanding, particularly in the $e^+e^-$ channel.

Initial results from radiative return analyses indicate that measurements
on the full data set from all experiments would produce a very interesting
cross-check on the LEP beam energy.

\section*{Acknowledgments}

I am grateful to the LEP 2 fermion working group and members of the
4 collaborations for their help in preparing this contribution.

\end{document}